\documentclass[aps,prl,showpacs,twocolumn]{revtex4}
\usepackage{amsmath, amsfonts, amsthm, amssymb, graphicx}
\usepackage{slashed}
\newcommand{\nc}{\newcommand}
\nc{\rnc}{\renewcommand}
\rnc{\d}{\mathrm{d}}
\nc{\D}{\partial}
\rnc{\t}{\tau}
\nc{\K}{\kappa}
\nc{\g}{\gamma}
\nc{\lrarrow}{\leftrightarrow}
\nc{\rg}{\sqrt{g}}
\rnc{\[}{\begin{equation}}
\rnc{\]}{\end{equation}}
\nc{\bea}{\begin{eqnarray}}
\nc{\eea}{\end{eqnarray}}
\nc{\nn}{\nonumber}
\rnc{\(}{\left(}
\rnc{\)}{\right)}
\nc{\dq}{\frac{\d^3 q}{(2\pi)^3}}
\rnc{\a}{\hat{a}}
\nc{\ep}{\epsilon}
\rnc{\tt}{\rightarrow}
\rnc{\inf}{\infty}
\rnc{\Re}{\mathrm{Re}}
\rnc{\Im}{\mathrm{Im}}
\nc{\ie}{{\it i.e.~}}
\nc{\iec}{{\it i.e.,~}}
\nc{\ta}{\bar{a}}
\nc{\vphi}{\varphi}      % use for cosmological scalar field
\nc{\tphi}{\bar{\vphi}}  % use for domain wall scalar field
\nc{\bphi}{\bar{\Phi}}   % use for domain wall perturbation
\nc{\tq}{\bar{q}}
\nc{\tK}{\bar{\K}}
\nc{\tV}{\bar{V}}
\nc{\tr}{\bar{r}}
\rnc{\ta}{\bar{a}}
\nc{\tH}{\bar{H}}
\nc{\tW}{\bar{W}}
\nc{\z}{\zeta}
\nc{\Z}{\mathcal{Z}}
\nc{\W}{\mathcal{W}}
\rnc{\H}{\mathcal{H}}
\nc{\<}{\langle}
\rnc{\>}{\rangle}
\rnc{\O}{\mathcal{O}}
\nc{\fnl}{f_{NL}}
\nc{\fnleq}{f_{NL}^{equil}}
\nc{\fnlloc}{f_{NL}^{local}}
\nc{\Lie}{\pounds}
\nc{\half}{\frac{1}{2}}
\nc{\bOmega}{\bar{\Omega}}
\nc{\bLambda}{\bar{\Lambda}}
\nc{\dN}{\delta N}
\nc{\gYM}{g_{\mathrm{YM}}}
\nc{\geff}{g_{\mathrm{eff}}}

\nc{\bN}{\bar{N}}
\nc{\bq}{\bar{q}}
\nc{\vbq}{\vec{\bar{q}}}

\begin{document}

\title{Observational signatures of holographic models of inflation}

\author{Paul McFadden${}^1$}
\email[]{P.L.McFadden@uva.nl}
\author{Kostas Skenderis${}^{1,2,3}$}
\email[]{K.Skenderis@uva.nl}

\affiliation{${}^1$ Institute for Theoretical Physics,
${}^2$ Gravitation and Astro-Particle Physics Amsterdam,
${}^3$ Korteweg-de Vries Institute for Mathematics,\\
Science Park 904, 1090 GL Amsterdam, the Netherlands.}

\date{1st October 2010}

\begin{abstract}

We discuss the phenomenology of recently proposed holographic
models of inflation, in which the very early universe is non-geometric
and is
described by a dual three-dimensional quantum field theory (QFT). We analyze
models determined by a specific class of dual QFTs
%has a field content
%consisting of massless fields (scalars, fermions and gauge fields),
%has a dimensionful coupling constant and
%is a super-renormalizable QFT
%with field content consisting of massless scalar, fermion and gauge fields and
%admitting a large $N$ 't Hooft limit
and show that they have the
following universal properties: (i)
they have a nearly scale invariant spectrum of small amplitude
primordial fluctuations, (ii) the scalar
spectral index runs as $\alpha_s = - (n_s-1)$,
%and a similar running for the tensors,
(iii) the three-point function
of primordial scalar perturbations is
of exactly the factorisable equilateral form with
$\fnleq=5/36$. These properties hold irrespective of the
details (e.g. field content, strength of interactions etc.) of the dual QFT
within the class of theories we analyze.
The ratio of tensors-to-scalars
is determined by the field content of the dual QFT and does not
satisfy the slow-roll consistency relations. Observations
from the Planck satellite should be able to confirm or exclude these models.

\end{abstract}

%\keywords{}

\maketitle

\paragraph{Introduction.}

A wealth of observational data has in recent years %over the recent past has 
transformed cosmology into a precise quantitative science, while new %and new %forthcoming %new 
observations 
%in the near future 
promise an unprecedented era of precision cosmology. 
In particular, the Planck satellite \cite{planck_website}, currently operational, is %expected to provide %release
%high resolution, high sensitivity maps of the
mapping the cosmic microwave background (CMB) anisotropies to an unprecedented degree of sensitivity and angular resolution.
%anisotropies in the very near future.
%
%that will severely tighten existing constraints on fundamental cosmological parameters.
%The reuslts will greatly improve our understanding of primordial cosmological perturbations
%
%in the very near future.
%
%In particular, the high resolution, high sensitivity observations of the cosmic microwave background %anisotropies
%produced by the Planck satellite should
%expected from the imminentn release of data from the Planck satellite
%
%In particular, the imminent
%arrival of data from the Planck satellite \cite{planck} will provide %is expected to provide %will provide %with the resulting 
%high sensitivity, high angular resolution maps of the microwave sky. The maps will
The results, expected shortly, will
severely tighten existing constraints on fundamental cosmological
parameters, and promise 
to greatly improve our understanding
of the primordial inhomogeneities that set the initial conditions %for the subsequent formation of structure.
for structure formation in our universe.
%primordial cosmological perturbations.  
Above all, it is hoped
that the results will shed light on the dynamical mechanism through which
these primordial cosmological perturbations were generated. 
%and as such  
%they presents a unique window to Planck-scale physics and a challenge and
%opportunity for fundamental theory. 

The leading candidate for this mechanism is %provided by 
the theory of inflation, which postulates that 
the early universe underwent a brief burst %period 
of accelerated expansion.
%In conventional inflation the underlying theory
%is gravity coupled to matter and the
%gravitational field equations admit as a solution
%an accelerating FRW specetime. 
The simplest models comprise a single scalar field, the inflaton,
coupled to gravity and equipped with a 
%In the simplest model, the
%matter part is a single scalar field, the inflaton, with 
potential function such that the scalar field
rolls slowly down to its minimum. 
%The theory of inflation
%provides an explanation for the initial
%inhomogeneities that set the initial conditions for
%structure formation in our universe. 
In the inflationary paradigm, one starts with a perturbative
quantization of fluctuations around the background spacetime
described by
%which corresponds to 
an accelerating Friedmann-Robertson-Walker (FRW) metric. %solution. 
As the universe
inflates, quantum fluctuations produced at very early times %the very early universe 
grow to superhorizon
scales and become classical. They then re-enter the
horizon in the post-inflationary era 
%and form the seed
%for structure formation. These primordial fluctuation
%have left 
%and leave the imprint in the CMB that is directly observed today.
and leave their imprint in the cosmic microwave background, %(CMB)
as directly observed today.
%that is directly observed today. 
Different matter content and interactions (potential) lead
to a variety of different models with different observational
signatures.

An underlying assumption in these scenarios is that
it is a valid approximation to use such a perturbative quantization
around the background FRW solution in the very early universe.
FRW solutions have a curvature singularity at early times, %however,
and, moreover, four-dimensional gravity is not UV complete
(it is a non-renormalizable theory). %For these reasons there is
%currently a lot of effort to embed inflation in string theory.
Although in some models
the perturbative approximation may be justified, in general
the quantum theory of the very early universe may be
strongly coupled with no useful perturbative gravitational
description. In this Letter we discuss models that
describe a universe of this type.

Over the last fifteen years a new principle has
emerged about the nature of any quantum theory of gravity,
namely that it should be holographic. In the context
of four-dimensional gravity, this means that there should
be an underlying complete description in terms of a
three-dimensional QFT without gravity. Concrete holographic
models were found in string theory, and, in these models,
the holographic correspondence is a
strong-weak coupling duality; namely, when the gravitational description is
weakly coupled, the dual QFT is strongly coupled and vice versa.

In our models, we will use perturbative three-dimensional QFT to model
a putative strongly coupled non-geometric phase of the very early universe. At the end
of this epoch, which is the analogue of the conventional inflationary epoch,
a weakly coupled geometric description becomes valid and we end up with a specific accelerating FRW
spacetime plus a specific set of inhomogeneities.
These inhomogeneities are not however linked with a perturbative quantization around the
FRW spacetime, but rather, they originate from the dynamics of the dual QFT.
This phase should then be matched to conventional hot big bang cosmology.
In this paper we will focus only on the phase before the hot big bang.
When we refer to late times this means the end of the ``holographic
inflationary epoch".

\paragraph{Holographic dualities.} The best understood examples of holographic
dualities are the ones obtained by considering brane configurations in string theory
and taking appropriate decoupling limits \cite{Maldacena:1997re}. In recent years, however, the use
of holographic methods has been greatly extended by taking a broader
perspective and using holography as a model of strong coupling physics.
%(even in cases where the applicability of these methods is not well  understood).
For example, gravitational computations in anti-de Sitter (AdS) spacetime %AdS 
were used to model the QCD
quark-gluon plasma, and, more recently, similar methods were applied to
condensed matter problems.

Here our perspective is similar:
%, although instead of using classical gravity to model strongly coupled QFT,
%as in these examples,
we will use perturbative quantum field theory to model
strong coupling gravitational dynamics. Furthermore,
as in the above examples, our emphasis will be on the phenomenology
of the models rather than the theoretical underpinnings.
As we will see, our holographic models are consistent with current
observations, yet their predictions are different to those of conventional inflation.
More importantly, near future observation may verify or exclude them,
so  the success or failure of these models will ultimately be %judged
measured 
by comparing with observational data.
If these models are confirmed by future observations, this would mark
a spectacular experimental verification of the idea of holography.

The holographic dualities which are best understood involve spacetimes
with negative cosmological constant. For the application at hand,
however, we would like to use such a duality for spacetimes
that are either de-Sitter or accelerating power-law spacetimes
at late times (recall that late times here mean the end of
the inflationary epoch). In recent work \cite{us1},
we proposed that such a duality may be obtained from
standard gauge/gravity duality by means of a specific
analytic continuation. The idea is sketched in Figure 1.

\begin{figure}[tr]
\includegraphics[width=8.5cm]{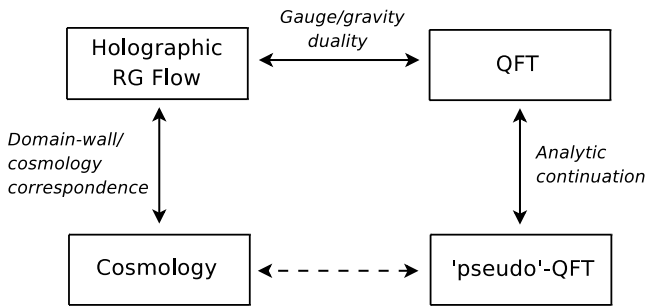}
\caption{The QFT dual to inflationary cosmology is operationally defined
using the correspondence of cosmologies to domain-walls and standard gauge/gravity duality.}
\end{figure}

In the first step (vertical left-hand side of Fig.~1), one uses the
fact that there is a one-to-one
correspondence between FRW cosmologies and domain-wall spacetimes
\cite{Cvetic:1994ya}. This can be understood as a
specific analytic continuation \cite{Skenderis:2006fb} that maps FRW spacetimes
that at late times approach either de-Sitter or power-law cosmologies,
to domain-wall spacetimes that are either asymptotically AdS or else have this
property in a specific conformal frame
\cite{Boonstra:1998mp}. These spacetimes
describe holographic RG flows and there is a
well-developed holographic set-up for them \cite{Skenderis:2002wp, Kanitscheider:2008kd}
(upper horizontal line of Fig.~1). Finally,
we express the domain-wall/cosmology analytic continuation
in terms of QFT variables (vertical right-hand side of Fig.~1). This amounts to a specific analytic continuation of
parameters and momenta that we
will discuss in more detail below.  Altogether, this gives a way to
relate a three-dimensional QFT with a corresponding cosmology.

The main piece of evidence for this holographic duality is that it correctly reproduces
conventional inflationary predictions in their regime of applicability, provided
standard gauge/gravity duality holds ({\it i.e.}, the upper horizontal line in Fig.~1 holds).
In conventional inflation,
the gravitational description is perturbative and thus one should find that inflationary
observables can be expressed in terms of (a specific analytic continuation of) strongly coupled correlators of three-dimensional QFT. Indeed, we showed in \cite{us1} that the power spectra may be expressed
in terms of 2-point functions of the stress-energy tensor, where the latter are computed gravitationally
using standard gauge/gravity duality, and the same also holds for the bispectrum \cite{us4}.

\paragraph{Holographic dictionary.} The holographic dictionary relates bulk and
boundary observables. In particular, bulk fields
are related to composite operators of the dual QFT. A prime example is the
correspondence between the bulk metric and the stress-energy tensor of the dual QFT.
The duality provides a prescription
for obtaining quantities in one theory by doing a computation in the other.
In our context, we would like to have holographic formulae relating cosmological
observables to correlation functions of the dual QFT. Such holographic formulae
were derived for the cosmological power spectrum in \cite{us1}, and results for the scalar bispectrum 
will appear in \cite{us4}. Here, we summarize these results.

The input for the holographic formulae are correlation functions of the
stress-energy tensor of the dual QFT. For the power spectra and the
bispectrum we only need the 2- and 3-point functions.  In momentum
space, the general form of the 2-point function is
\[
\label{ABdef}
\<T_{ij}(\tq)T_{kl}(-\tq)\> = A(\tq,\bar{N})\Pi_{ijkl}+B(\tq,\bar{N})\pi_{ij}\pi_{kl},
\]
where $\bar{q}$ is the magnitude of the 3-momentum and
we indicate explicitly that the
QFT coefficients depend on $\bar{N}$, the rank of the
gauge group of the dual QFT. %(the QFT admit a 't Hooft large $\bar{N}$ limit).
We suppress the dependence on other parameters (coupling constants, {\it etc}.).
$\Pi_{ijkl}$ is the three-dimensional transverse traceless projection operator,
$
\Pi_{ijkl}= (\pi_{ik}\pi_{jl}+\pi_{il}\pi_{jk}- \pi_{ij}\pi_{kl})/2$,
and
$
\pi_{ij} = \delta_{ij} - \bar{q}_i \bar{q}_j/\bar{q}^2.
$

The cosmological scalar and tensor power spectra are then linked to the
2-point function via
\[
\label{result}
\Delta^2_S(q) = \frac{-q^3/16 \pi^2}{ \Im B(-i q,-i N)}, \
\Delta^2_T(q) = \frac{-2 q^3/\pi^2}{ \Im A(-i q,-iN)},
\]
{\it i.e.}, we analytically continue the magnitude of the
3-momentum and the rank of the gauge group as
\[
\label{QFT_cont}
 \bar{N} = -iN, \qquad \bar{q}=-iq.
\]

The holographic formula for the bispectrum of curvature perturbations
is given by \cite{us4}
\begin{align}
\label{Main_result2}
&\<\!\<\hat{\z}(q_1)\hat{\z}(q_2)\hat{\z}(q_3)\>\!\>
=-\frac{1}{4} \frac{1}{\prod_i \Im \<\!\<T(\tq_i)T(-\tq_i)\>\!\>} \nn \\
& \qquad\qquad \cdot\Im \big[\<\!\<T(\tq_1)T(\tq_2)T(\tq_3)\>\!\> +
\sum_i \<\!\<T(\tq_i)T(-\tq_i)\>\!\> \nn \\
&\qquad\qquad
-2\big( \<\!\<T(\tq_1)\Upsilon(\tq_2,\tq_3)\>\!\>{+}\mathrm{cycl.\,perms}\big)
\big],
\end{align}
where $T=\delta^{ij} T_{ij}$ and on the r.h.s.~one should use (\ref{QFT_cont}) before taking the imaginary part. The (Fourier transform of the)
operator $\Upsilon(\tq_1,\tq_2)$ is defined by
$
\Upsilon(x_1,x_2) = \delta^{ij}\delta^{kl}
\delta T_{ij}(x_1)/\delta g^{kl}(x_2)|_{g_{mn}=\delta_{mn}},
$
{\it i.e.}, one first obtains the stress-energy tensor $T_{ij}(g^{kl})$
of the theory coupled to gravity, differentiates w.r.t.~$g^{kl}$, and then
sets the background metric equal to the flat metric.
The double bracket notation indicates that we have extracted
the momentum-conserving delta function (times $(2 \pi)^3$), {\it e.g.}, %. For example,
$\<T(\vbq_1)T(\vbq_2)T(\vbq_3)\> = (2\pi)^3\delta(\sum_i\vbq_i)\<\!\<T(\tq_1)T(\tq_2)T(\tq_3\>\!\>$.

As mentioned above, if
we are in the regime where gravity is weakly coupled, we may use
standard gauge/gravity duality to holographically compute the QFT correlators:
the holographic formulae above then reproduce standard inflationary results.
In this Letter we are interested in the opposite regime, where gravity is
strongly coupled at early times, and we will therefore use perturbative QFT
methods to compute the QFT correlation functions.

\paragraph{The model.} Ideally, the precise form of the
QFT model should be derived from first principles. In the absence of
such a derivation, we proceed by using as few assumptions as
possible and aim at obtaining universal results that do not
depend on the details of the dual QFT. The QFTs that currently
have holographic duals are either theories that in the UV 
become conformal or QFTs with a generalized conformal structure
\cite{Jevicki:1998ub,Kanitscheider:2008kd}. Here, we focus on the latter %second
class of theories.
An example of such a theory
is the maximally supersymmetric super-Yang-Mills
theory in three dimensions. In such theories, all terms in the Lagrangian
scale in the same way, but their dimension is not equal to the
spacetime dimension. To be able to use perturbative QFT methods
we require standard kinetic terms, so the action is of the form
\begin{align}
\label{Lfree}
S =\frac{1}{g_{\mathrm{YM}}^2}\int \d^3 x\,
 \mathrm{tr}\Big[&\frac{1}{2} F^I_{ij}F^I_{ij} +  \frac{1}{2} (\D\phi^J)^2
+  \frac{1}{2} (\D\chi^K)^2\nn \\
&
+ \bar{\psi}^L \slashed{\D} \psi^L + \mathrm{interactions}\, \Big],
\end{align}
where all fields are massless and we have
$\mathcal{N}_A$ gauge fields $A^I \ (I{=}1, \dots,
\mathcal{N}_A)$, $\ \mathcal{N}_\phi$ minimally coupled scalars $\phi^J\ (J{=}1,
\dots, \mathcal{N}_\phi)$, $\ \mathcal{N}_\chi$ conformally coupled scalars
$\chi^K\ (K{=}1,\dots, \mathcal{N}_\chi)$ and $\mathcal{N}_\psi$
fermions $\psi^L \ (L{=}1,\dots, \mathcal{N}_\psi)$.
Note that $g_{\mathrm{YM}}^2$ has dimension one in three dimensions.
The trace is over the gauge indices and for concreteness we
consider the gauge group to be $SU(\bar{N})$ and all fields to
transform in the adjoint of $SU(\bar{N})$. Any other choices that would admit a large $\bar{N}$
limit would be as good. One can write down the general form of the
interactions such that all terms have dimension four, but in fact (almost)
all of the results to follow are independent of the specific form
of interactions. QFTs of this form are super-renormalizable and the  
dimensionful coupling constant acts as an infra-red cut-off \cite{Jackiw:1980kv}, so all
computations are under control.

QFTs with a generalized conformal structure are dual to domain-walls with asymptotic
power-law scaling. Thus, at the end of the holographic inflationary epoch, Einstein gravity
becomes a good description again and the universe is now described by
a power-law accelerating spacetime
\[ \label{power_law}
\d s^2 = -\d t^2 + t^{2n} \d\vec{x}\,^2, \quad n>1,
\]
with a specific set of inhomogeneities that follow from the
holographic computation to be discussed below. Note that
(\ref{power_law}) is only valid towards the end of this epoch and is
the metric that should be matched to the subsequent evolution. Let us
emphasize that conventional inflationary theory based on
(\ref{power_law}) is disfavoured by current observations \cite{Komatsu:2008hk}.
%For example, the corresponding power spectrum is not scale invariant.
The inhomogeneities that we discuss below are not however linked with perturbation
theory around (\ref{power_law}), but rather, with the dynamics of
(\ref{Lfree}). We will leave $n$ as a free parameter.
We note, however, that the case $n=7$ is distinguished
because the corresponding domain-wall solution is the near-horizon
limit of the D2-brane solution.

\paragraph{Predictions.} We are now ready to present the predictions of this model.
For this, we need to compute the 2- and 3-point functions
of the stress-energy tensor. The details of the computation of the 2-point function
has been reported in \cite{us1} and for the 3-point function
will appear in \cite{us4}. Here, we provide an overview of these results and discuss their
implications.

The results for the 2-point function up to 2-loops is
\begin{align}
\label{2-loop_result}
 A(\bar{q},\bar{N}) &= C_A \bar{N}^2 \bar{q}^3[1+D_A g_{\mathrm{eff}}^2 \ln (\bar{q}/\bar{q}_0) +O(g_{\mathrm{eff}}^4)], \nn \\
B(\bar{q},\bar{N}) &= C_B \bar{N}^2 \bar{q}^3[1+D_B g_{\mathrm{eff}}^2 \ln (\bar{q}/\bar{q}_0)+O(g_{\mathrm{eff}}^4)], 
\end{align}
where
$
C_A =(\mathcal{N}_A + \mathcal{N}_\phi +\mathcal{N}_\chi+2\mathcal{N}_\psi)/256, \
C_B = (\mathcal{N}_A+\mathcal{N}_\phi)/256$
and $g_{\mathrm{eff}}^2=g_{YM}^2 \bar{N}/\bar{q}$ is the dimensionless effective
coupling constant. 
The renormalization scale $\bar{q}_0$ may be identified with the cosmological pivot scale $q_0$ 
via (\ref{QFT_cont}).
$D_A$ and $D_B$ are numerical coefficients of order one
whose value depends on the field content and the precise form
of the interactions. To compute $D_A$ and $D_B$ precisely requires summing
all the relevant 2-loop diagrams. Notice that the effective coupling
remains real under the continuation (\ref{QFT_cont}). 
%and hence the 2-point function has a well-defined continuation. 
%The logarithmic terms originate from
%potential infra-red divergences, but as discussed in \cite{Jackiw:1980kv},
%the dimensionful coupling constant effectively acts as an infra-red cut-off.
Inserting now in (\ref{result}) and using the standard
cosmological parameterizations,
$
 \Delta^2_S(q) = \Delta^2_S(q_0) (q/q_0)^{n_S(q)-1},\ \Delta^2_T(q) =
\Delta^2_T(q_0) (q/q_0)^{n_T(q)}$,
where $\Delta^2_{S/T}(q_0)$ is the scalar/tensor amplitude at some chosen
pivot scale $q_0$, and $n_{S/T}(q)$ is the scalar/tensor spectral tilt,
we find for the amplitudes
\[
 \Delta_S^2(q_0) = \frac{1}{16\pi^2 N^2 C_B}, \ 
\Delta_T^2(q_0) = \frac{2}{\pi^2 N^2 C_A},
\]
and for the indices
\[
n_S(q){-}1 = {-}D_B g_{\mathrm{eff}}^2 + O(g_{\mathrm{eff}}^4), \quad
n_T(q) = {-}D_A g_{\mathrm{eff}}^2 + O(g_{\mathrm{eff}}^4).
\]
 From the WMAP data \cite{Komatsu:2008hk} we have $\Delta^2_S(q_0) \,{\sim}\, O(10^{-9})$, hence $N \,{\sim}\, O(10^4)$, justifying our use of the large $N$ limit.
Furthermore, $(n_s{-}1) \,{\sim}\, O(10^{-2})$ at $q_0\,{=}\,0.002\,
\mathrm{Mpc}^{-1}$, and hence $g_{\mathrm{eff}}^2(q_0) \,{\sim}\, O(10^{-2})$
also, justifying our perturbative treatment of the QFT.
(Even at the momentum scale $q_h \,{\sim}\, 2\times 10^{-4} \mathrm{Mpc}^{-1}$
corresponding to the present comoving horizon radius $r_h \,{\sim}\, 14 \mathrm{Gpc}$,
$g_{\mathrm{eff}}^2(q_h) \,{\sim}\, O(0.1)$ hence all the comoving scales seen in the CMB today lie
within the weak-coupling regime where $g_{\mathrm{eff}}^2$ is small).
In other words, the two small numbers that appear in the data,
the amplitude of the primordial fluctuations and the deviation
from scale invariance, appear rather naturally in the dual QFT.

%Note that momentum scales in the dual QFT correspond to comoving momenta in
%cosmology.  For a flat Lambda-CDM universe, the comoving horizon
%radius is $d_{h} \sim 14 GpC$ \cite{Komatsu:2008hk}, and so the smallest
%comoving momentum scale we could ever observe today is $q_h = 2\pi
%/(2d_h) \sim 2 \times 10^{-4} MpC^{-1}$.  Taking $g_{\mathrm{eff}}^2 \sim
%O(10^{-2})$ at the WMAP pivot scale of $q_0= 2\times 10^{-3}
%MpC^{-1}$, we find $g_{\mathrm{eff}}^2(q_h) \sim O(10^{-2}) \times (q_0/q_h) \sim
%O(0.1)$.  Thus, all comoving scales seen in the CMB today lie within
%the weak-coupling regime where $g_{\mathrm{eff}}^2$ is small.

\begin{figure}[tr]
\includegraphics[width=6.3cm]{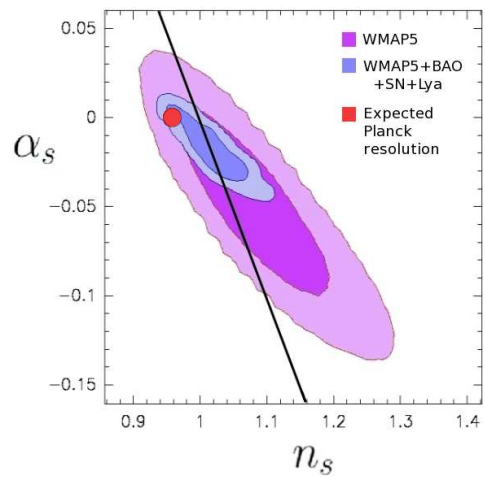}
\caption{\label{running_fig}
The straight line is the leading order prediction of our
holographic model for the
correlation of the running $\alpha_S$ and the scalar tilt $n_S$.
The data illustrate the $1\sigma$ and $2\sigma$ constraints (marginalizing over tensors)
at $q_0 = 0.002\, \mathrm{Mpc}^{-1}$, constructed using \cite{plotter}.
As a rough guide to future developments,
%an approximate comparison, 
the expected size of the $1\sigma$ allowed region for the Planck data is also shown (based on simulations reported in \cite{planck} with input
values $n_S{=}0.957$, $\alpha_S{=}0$ but excluding tensors). %is also shown \cite{planck}.
}
\end{figure}

The tensor-to-scalar ratio $r\,{=}\,\Delta_T^2/\Delta_S^2 \,{=}\, 32 C_B/C_A$ depends
on the field content of the dual QFT; note however that $r$ is not parametrically
suppressed as in slow-roll inflation, nor %of course
does it satisfy the conventional slow-roll consistency condition $r \,{=}\, -8n_T$. %\cite{liddle}.
To determine whether the spectral tilts are red or blue requires
evaluating the signs of $D_A$ and $D_B$, which in general will depend
on the field content and interactions of the QFT.
It is nonetheless still possible to
extract predictions which are independent of the details of the QFT: for
example, in these models, the scalar spectral index runs as
\[
 \alpha_S = \d n_S/\d\ln q =  -(n_S{-}1) + O(g_{\mathrm{eff}}^4).
\]
This prediction is qualitatively different from slow-roll inflation,
for which $\alpha_S/(n_S{-}1)$ is of first-order in slow-roll
\cite{Kosowsky:1995aa}.  Running of this form is still consistent with the WMAP
observational constraints for a wide range of values of $n_S$ and
$\alpha_S$ \cite{Komatsu:2008hk}, as illustrated in Fig.~\ref{running_fig}. In the near future,
observations from the Planck satellite will determine $n_S$ and
$\alpha_S$ to such accuracy that this measurement alone could exclude these models.
The expected $1\sigma$ uncertainty in $n_S$ and $\alpha_S$ of $\sim 0.005$ quoted in \cite{planck} is illustrated in  Fig.~\ref{running_fig}.

We now move to the discussion of the bispectrum. To 1-loop order we obtain
\begin{align}
&\<\!\<T(\bq_1)T(\bq_2)T(\bq_3)\>\!\> + \sum_i \<\!\<T(\bq_i)T(-\bq_i)\>\!\> \nn \\ 
&\quad -2\big(\<\!\<T(\bq_1)\Upsilon(\bq_2,\bq_3)\>\!\> +\mathrm{cyclic\,perms}\big)\nn\\
&=2 C_B \bN^2\big(2\bq_1\bq_2\bq_3+\sum_i \bq_i^3-(\bq_1\bq_2^2+5\,\mathrm{perms})\big). 
\end{align}
Inserting in (\ref{Main_result2}) we find
\begin{align}
\label{template}
&\<\!\<\hat{\z}(q_1)\hat{\z}(q_2)\hat{\z}(q_3)\>\!\> =
6\mathcal{A}^2 \fnleq\Big(\frac{3}{5}\Big)  \Big(\prod_i q_i^{-3}\Big)\\
& \qquad\cdot \big({-}2q_1q_2q_3-\sum_i q_i^3+(q_1q_2^2+5\,\mathrm{perms})\big), \nn
\end{align}
with
$
 \fnleq = 5/36
$
(in the sign convention of \cite{Komatsu_review})
and $\mathcal{A} = 2\pi^2 \Delta_S^2(q)$. %q^3\<\!\<\hat{\z}(q)\hat{\z}(-q)\>\!\>$.
%(equal to $A=1/(8 C_B N^2)$ in our case).
Interestingly, this is {\it exactly} equal to the factorisable equilateral
template introduced in \cite{Creminelli:2005hu}. Note that all dependence on the 
field content cancels out. 

%While an $\fnleq$ of this magnitude is probably too small to be detected by Planck directly \cite{Komatsu:2001rj},

\paragraph{Conclusions.}

%Holography enables the description of a universe that was non-geometric at early times,
%and leads to distinctive observational signatures.

% We presented a holographic description of a universe that was non-geometric
% at early times and discussed possible observational signatures.
% %
% Equipped with precise predictions for the running of the spectral
% index and for the form and magnitude of the expected primordial
% non-Gaussianity, the holographic models described here stand ready to
% confront the forthcoming observational data.  
% %
% The Planck data has the
% power to comfortably refute both predictions: alternatively, it may
% reveal the first observational signature of a primordial holographic
% era.

We presented a holographic description of a universe that was non-geometric
at early times and discussed possible observational signatures.
%
%The precise prediction for the bispectrum given above
While the specific non-Gaussianity predicted above is mostly likely too small to be detected by Planck directly \cite{Komatsu:2001rj},
the observation of a large $\fnl$ values would serve to refute these models.
The predicted running of the spectral index provides a more certain test of the holographic models described here,
and may potentially provide a first indication of the existence of a primordial holographic era.

\nopagebreak

{\it Acknowledgments.}  We thank NWO for support.

\end{document}